\documentclass{iau} 
\usepackage{graphicx}
\usepackage{amsmath}

\newcommand{\HII}{H{\sc ii}}
\newcommand{\sunn}{$_{\odot}$}

\title[DDO68-V1: an extremely metal-poor LBV in a void galaxy] 
{DDO68-V1: an extremely metal-poor LBV in a void galaxy}

\author[Yulia Perepelitsyna \& Simon Pustilnik]
   {Yulia Perepelitsyna$^1$  \and Simon Pustilnik$^1$}

\affiliation{$^1$Special Astrophysical Observatory Russian Academy of Sciences, \\ 
Nizhnij Arkhyz, Karachai-Circessia, 369167, Russia \\ email: {\tt jlyamina@yandex.ru, sap@sao.ru}}

\pubyear{2018}
\volume{344}  
\jname{Dwarf Galaxies: From the Deep Universe to the Present}
 \editors{K.~McQuinn, S.~Stierwalt, eds.}
\begin{document}

\maketitle

\begin{abstract}
The lowest metallicity massive stars in the Local Universe with 
$Z \sim$($Z$\sunn/50-$Z$\sunn/30) are the crucial objects to 
test the validity of assumptions in the modern models of very 
low-metallicity massive star evolution. These models, in turn, 
have major implications for our understanding of galaxy and 
massive star formation in the early epochs. DDO68-V1 in a void 
galaxy DDO68 is a unique extremely metal-poor massive star.
Discovered by us in 2008 in the HII region Knot3 with 
$Z = Z$\sunn/35 [12+$\log$(O/H)$\sim$7.14], DDO68-V1 was 
identified as an LBV star. 
We present here the LBV lightcurve in V band, combining own new 
data and the last archive and/or literature data on the light of 
Knot3 over the 30 years. We find that during the years 2008-2011 
the LBV have experienced a very rare event of `giant eruption' 
with V-band amplitude of 4.5 mag ($V \sim24.5^m -20^m$).

\end{abstract}

\firstsection 
\section{Introduction}

Luminous Blue Variable (LBV) stars represent a short (about
or less than 0.1 Myr) transient phase of massive star evolution
from the main sequence hydrogen burning O stars to the
core-helium burning Wolf-Raye (WR) stars.

Evolution of massive stars with the lowest known metallicities
is crucial for understanding the early galaxy formation and
evolution at high redshifts due to their great energy
release/feedback (e.g., \cite[Barkana \& Loeb (2001)]{Barkana01}).

The most metal-poor {\bf massive} stars are currently identified
in several extremely metal-poor ($Z \sim Z$\sunn/45-- $Z$\sunn/35)
dwarf galaxies. Most of these extreme galaxies are found in nearby voids.

Stellar evolution models (including those with the fast rotation)
have substantially advanced during the last decade. However,
the direct comparison of the model predictions with the properties
of real extremely metal-poor massive stars is still absent.
Such studies should await for the next generation extremely
large telescopes.

\section{Overview}

DDO68, at the distance D=12.75 Mpc, is one of the most metal-poor
galaxies ($Z \sim Z$\sunn/35) residing in the nearby Lynx-Cancer
void. DDO68 is a merger of low-mass gas-rich components
(\cite[Ekta, Chengalur, Pustilnik (2008)]{Ekta08},
\cite[Makarov et al. (2017)]{Makarov17}). Its very low-Z gas
was identified with BTA spectra in 2005.
Most of SF regions are found at the periphery, mainly in the
`Northern ring' and the `Southern tail'
(\cite[Pustilnik, Kniazev, \& Pramskij (2005)]{DDO68},
\cite[Izotov \& Thuan (2007)]{IT07}).

\begin{figure}[h]
	\begin{center}
  \includegraphics[width=5.3in]{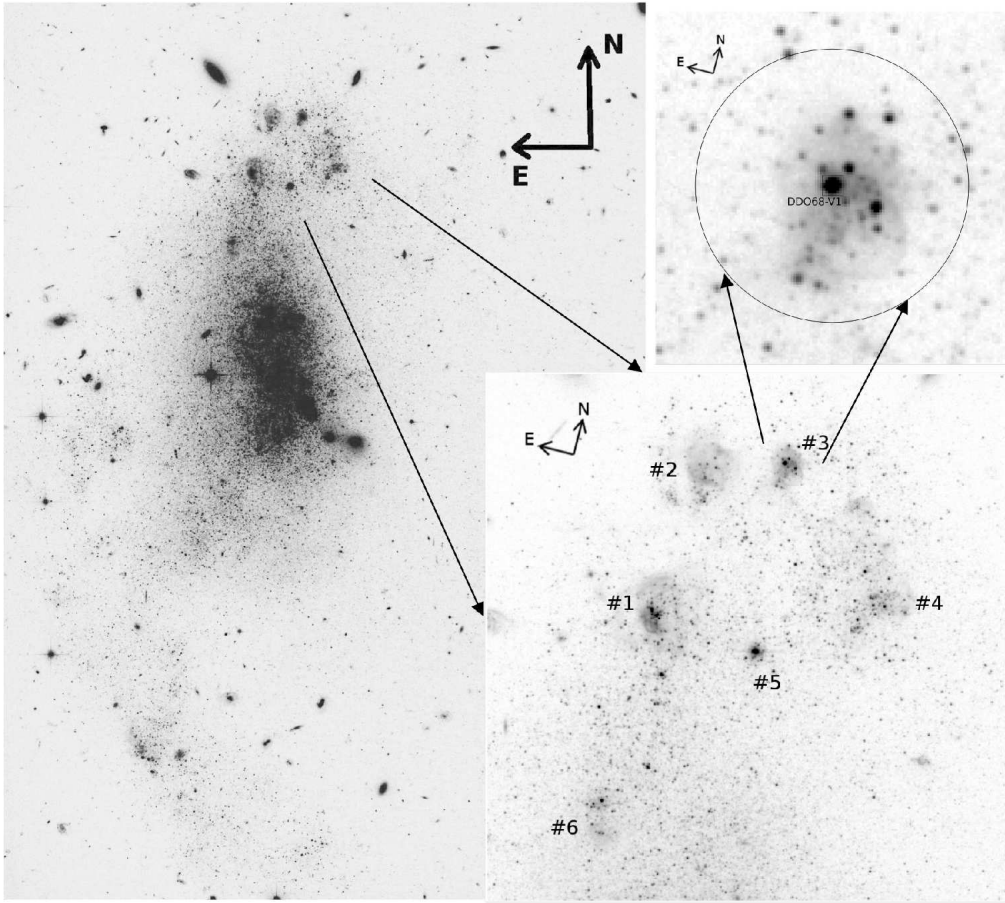}
  \caption{
The part of the HST image of DDO68 in $W606$ ($V$) band
centered on the region Knot 3 with the used aperture superimposed
($D_{\rm aper} =$ 5$''$).  DDO68-V1 is in the center of the aperture.
    }
    \label{fig1}
	\end{center}
\end{figure}

In 2008 we discovered in its SF Knot 3 (Fig.~\ref{fig1})
a transient which was identified with an LBV
(\cite[Pustilnik et al. (2008)]{LBV}, see also
\cite[Izotov \& Thuan (2009)]{Izotov09}). Hubble Space Telescope
(HST) images of DDO68 were obtained in May 2010 with $ACS$
for Proposal GO~11578 (PI~A.Aloisi) and presented in papers
by \cite[Sacchi et al. (2016)]{Sacchi2016} and by
\cite[Makarov et al. (2017)]{Makarov17}.

\begin{figure}[h!]
  \begin{center}
  \includegraphics[angle=-90,width=8.0cm,clip=]{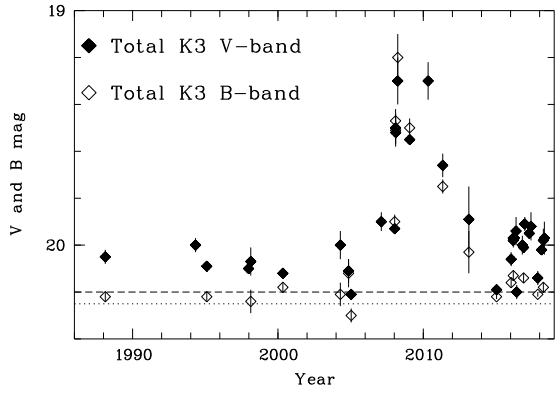}
   \caption{
The lightcurve of Knot~3 in DDO68 in $V$ and $B$ bands since
1988 based on new and archive data and the data from
\cite[Pustilnik et al. 2017]{DDO68LBV}. All magnitudes are
for the aperture with r = 2.5$''$.
      }
      \label{fig2}
   \end{center}
\end{figure}

The lightcurve of Knot~3 (Fig.~\ref{fig2}) in DDO68 in
$V$ and $B$ bands since 1988 is based on the new and archive data
and the data from \cite[Pustilnik et al. (2017)]{DDO68LBV}.
All magnitudes are for the aperture with r=2.5$''$. The dotted
lines at
$V = 20.20$ and $B = 20.25$ correspond to the minimal observed
light of the entire Knot~3. These minimal levels were slightly
reduced due to a more advanced background determination with respect
of that adopted in paper by
\cite[Perepelitsyna \& Pustilnik (2017)]{PerepelPustilnik2017}.
These magnitudes are consistent,
in particular, with  Knot~3 light on the night 2005.01.12, when
the LBV was too faint and did not show up in the spectrum of Knot~3.

\begin{figure}[h!]
   \begin{center}
    \includegraphics[angle=-90,width=6.5cm,clip=]{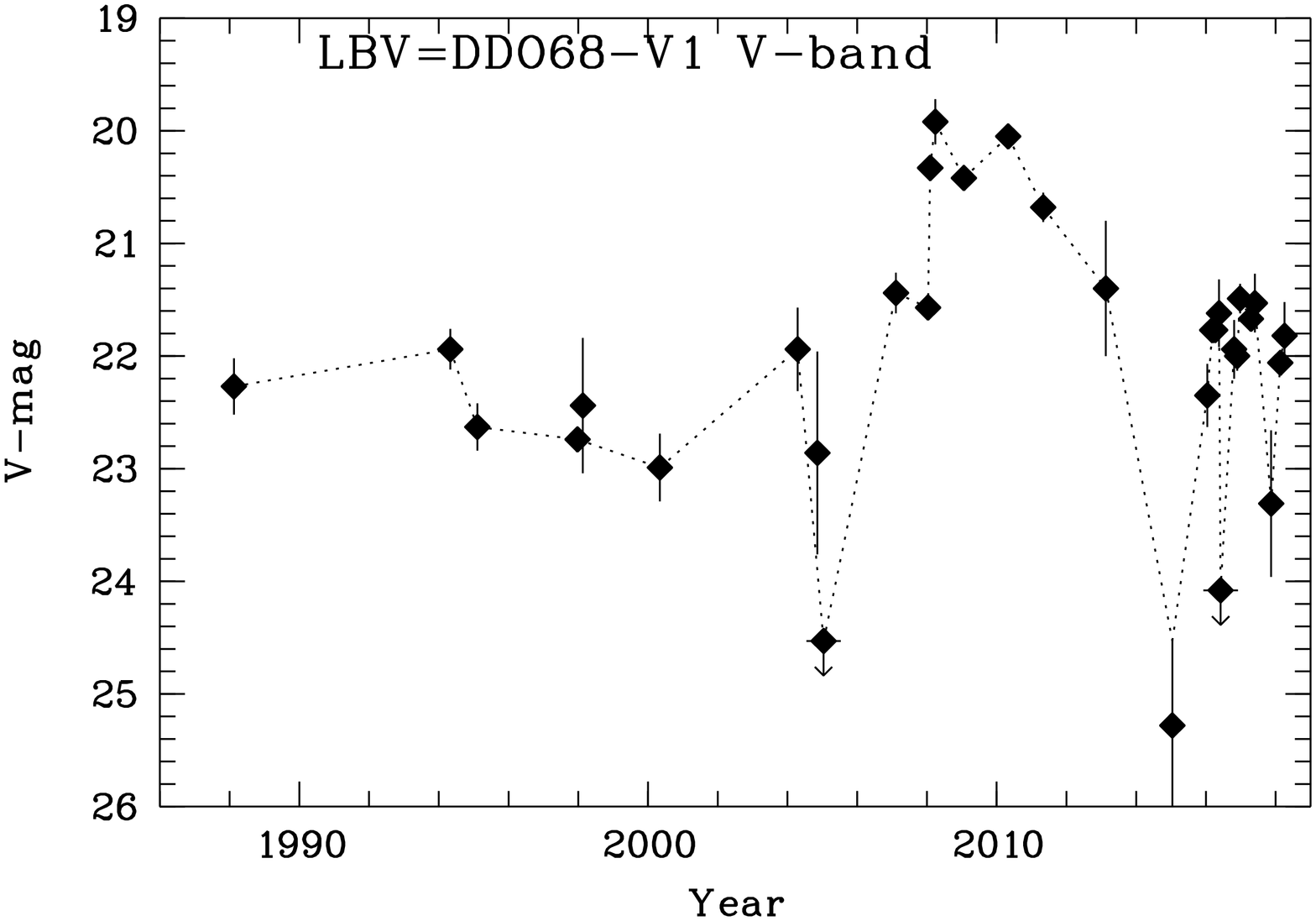}
    \includegraphics[angle=-90,width=6.5cm,clip=]{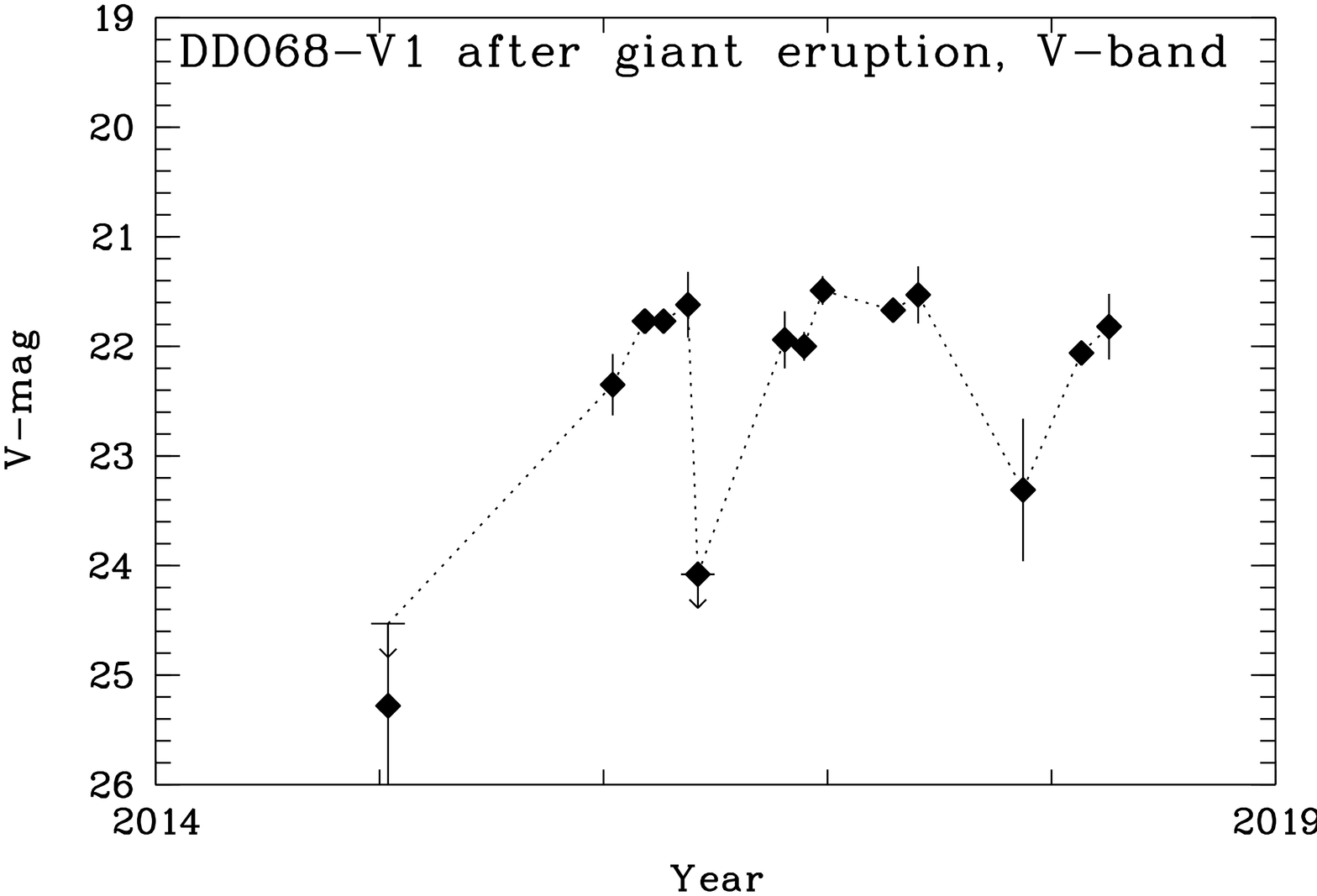}
    \caption{
 {\bf Left panel:} Light curve of the LBV in $V$-band (filled lozenges).
 With except of one direct photometry (the {\it HST} W606 image in May 2010),
all other magnitudes are derived as the `residual light' via the subtraction of
the constant luminosity of the underlying \HII\ region ($V = 20.20$) from the
previous lightcurve. Lozenges with arrows indicate 3$\sigma$ upper limits.
 {\bf Right panel:} Close-up of the LBV light curve in $V$-band
for period of 2015 -- 2018. There is an indication of
the phenomenon of S~Doradus type variations (\cite[Sterken, 2003]{SDor}).
      }
      \label{fig3}
  \end{center}
\end{figure}

With except of one direct photometry (the HST image), all other
magnitudes are derived as the `residual light' via subtraction
of the constant luminosity of the underlying HII region ($V=20.20$)
from the lightcurve on the Fig.~\ref{fig3}.
We observe a very rare case of LBV `giant eruption'
(\cite[Smith \& Owocki (2006)]{Smith2006}) during the years 2008-2011,
with  the total amplitude of the LBV optical variability
$\delta$V$\sim$4.5$^m$, reaching $M_{\rm V}$ = -10.5.
Series of `giant eruptions' in LBVs which form several expanding
shells, can precede their SN explosions at rather short time scale.
Observations of light variations of DDO68-V1 after the `giant eruption',
since Year 2015 reveal the behaviour resembling the  phenomenon
of S Doradus (\cite[Sterken, 2003]{SDor}). In the right panel
of Figure 3, the photometric variability is observed up to
2.5$^m$ over the periods of 0.5--2 years.

\section{Implications and conclusions}

\begin{enumerate}
\item We extend the recently published lightcurve for the period
of 2005 -- 2015 for DDO68-V1 (\cite[Pustilnik et al. 2017]{DDO68LBV}),
adding our fresh (years 2016--2018)
Zeiss-1000 and BTA telescopes photometry of the HII region Knot 3
(containing the LBV = V1) and the photometry from the archive
images at ten epochs with ten different telescopes over the period
of 1988 -- 2013.
\item The data allow us at the first time to determine the reliable
amplitude of this LBV lightcurve. All available data suggest that
the LBV $V$-band light varied during the last decade in the range
of $\sim$20.0$^m$ to fainter than 24.5$^m$. This corresponds to
the absolute magnitude $M_{\rm V}$ range of -6.0$^m$ to -10.5$^m$.
\item If the photometric behavior of the most metal-poor LBV is
similar to that of more typical LBVs, the DDO68-V1 light variations
during the last 28 years suggest that it underwent a `giant eruption'
during the years 2008 -- 2011.
\item We call to the community for the campaign of DDO68-V1
multiwavelength monitoring that can give the new insights in the lowest
metallicity LBV properties and prove the substantial increase of
its bolometric luminosity.
\item Having in mind other known examples of extragalactic SN impostors,
one can occasionally catch this unique object in the SN impostor
phase. Moreover, in the case of the great luck, we can catch even the
unique case of a nearby SNII explosion related to the extremely low-Z
massive star.
\end{enumerate}

\vspace{0.5cm}
The full-format paper presenting all details of observational data
and their analysis as well as a wider discussion of all available data
is prepared for publication in MNRAS.

\section*{Acknowledgements}
The work was supported by the grant of Russian Science Fund No. 14-12-00965.
The authors thank O.~Spiridonova, V.~Goransky and A.~Moskvitin for
their help with DDO68 observations at the SAO 1m telescope. The authors
are
grateful to L.~van~Zee, D.~Hunter, B.~Elmegreen, U.~Hopp, L.~Makarova,
R.~Swaters, B.~Mendez, V.~Taylor, R.~Jansen, R.A.~Windhorst, S.C.
Odewan, J.E.~Hibbard for providing archival CCD images of DDO68
obtained for their observational programs. We are pleased to thank
P.~Kaigorodov and D.~Kolomeitsev for their kind help in extracting
the data from archive tapes.

\end{document}